\documentclass[onecolumn]{emulateapj}

\begin{document}

\title{PHOTOMETRIC AND OBSCURATIONAL COMPLETENESS}
\author{Robert A.\ Brown}
\affil{Space Telescope Science Institute,\altaffilmark{1} 3700 San Martin Drive, Baltimore, MD 21218}
\email{rbrown@stsci.edu}
\altaffiltext{1}{The Space Telescope Science Institute is operated by the Association of Universities for Research in Astronomy, Inc.\ under NASA contract NAS 5-26555.}

\begin{abstract}
We report a method that uses ``completeness" to estimate the number of exrasolar planets discovered by an observing program with a direct-imaging instrument. We develop a completeness function for Earth-like planets on ``habitable" orbits for an instrument with a 
central field obscuration,  uniform sensitivity in an annular detection zone, and limiting sensitivity that is expressed as a ``delta magnitude" 
with respect to the star, determined by systematic effects (given adequate exposure time). We demonstrate our method of estimation by applying it to our understanding of the coronagraphic version of the \textit{Terrestrial Planet Finder} (\textit{TPF-C}) mission as of October 2004. We establish an initial relationship between the size, quality, and stability of the instrument's optics and its ability to meet mission science requirements. We provide options for increasing the fidelity and versatility of the models on which our method is based, and we discuss how the method could be extended to model the \textit{TPF-C} mission as a whole, to verify that its design can meet the science requirements. 
\end{abstract}

\keywords{planetary systems---techniques: high resolution---instrumentation: high angular resolution}

\section{INTRODUCTION}

We report a method that uses ``completeness'' to estimate the number of extrasolar planets discovered by an observing program with a direct-imaging instrument. The method is useful for optimizing the design of a mission and verifying that scientific requirements will be met.

We demonstrate our method for the case of detection in reflected starlight at visible wavelengths, particularly using the coronagraphic version of the \textit{Terrestrial Planet Finder} instrument (\textit{TPF-C}). Nevertheless, adaptation to the thermal infrared would be straightforward. The method requires simply (1)~probability distributions of the orbital elements and physical characteristics that  represent the planetary population of interest, (2)~algorithms for computing a planet's position in space from orbital elements, (3)~a pool of possible target stars, including their positions in space and physical characteristics, (4)~algorithms to compute a planet's specific flux from its physical characteristics, position in space, the observing geometry, and relevant stellar properties, (5)~algorithms to compute the signal-to-noise ratio (\textit{SNR}) of a detection from the planet's flux and image position on the focal plane, stellar and circumstellar properties, and the exposure time, and (6)~algorithms for choosing the next target star, defining its exposure time, and computing the full cost of an observation in terms of time.  

Our method uses the concept of photometric and obscurational completeness, which is the fraction of possible planets that will be detected for a given star, exposure time, and distribution of sensitivity on the focal plane (the ``detection zone''). Assuming the star has exactly one planet, the number of planets found (i.e., one or zero) is a Bernoulli random variable with expectation value 
($<$$n_\mathrm{p}$$>$) equal to the completeness. The expectation value of the total number of planets found by an observing program on multiple stars ($<$$\Sigma n_\mathrm{p}$$>$) is the sum of the completenesses of the individual stars. If the occurrence rate is smaller than unity, the expected yield of planets is proportionally smaller.

It is our understanding that \textit{TPF-C} begins a ``pre-phase A'' study in 2005, which will culminate in at least one strawman design that has been demonstrated to meet mission science requirements. The verification of that achievement will require simulating the mission to which the strawman design is referenced, or ``design-reference mission'' (DRM). We will discuss extensions of our estimation method that will be essential to perform those simulations and attain that verification.

For simple cases, a functional representation of completeness may be possible and convenient. This is true of our demonstration, in which the completeness of a single, initial searching observation is represented by a function of two independent variables: the apparent separation of planet and star and the planet-to-star flux ratio, expressed as a ``delta magnitude.'' For more complex cases---for example, detection zones with more detailed structure or multiple observing epochs---it will be necessary to implement the concept of completeness directly, by a ``Monte Carlo,'' computer orrery of possible planets, keeping track in detail of which possible planets have been discovered and which not as an observing program progresses.

In an earlier paper, we analyzed the selection effects of a circular central field obscuration on the completeness of searches for extrasolar planets by direct imaging (Brown 2004a). We introduced the term ``obscurational completeness'' to refer to the fraction of a population of possible planets that is detectable according to the obscurational criterion alone:
\begin{equation}
s>a_0~~, 
\end{equation}
where $s$ is the apparent separation between planet and star in AU, and $a_0$ is the projected radius of a central field obscuration in AU. We did not include photometric criteria for detection, which meant, on the one hand, that the results would be valid whether a detection technique used reflected stellar radiation at ultraviolet, visible, or near-infrared wavelengths or used thermal-infrared planetary radiation at mid- or far-infrared wavelengths. On the other hand, by implicitly assuming that an observation could detect any and all unobscured planets, obscurational completeness could only be an upper limit to practical completeness, which must take photometric errors and limitations into account. This paper redresses that shortcoming by analyzing ``photometric and obscurational completeness'' in visible light, taking into account the brightness of planets, which varies according to the inverse-square law of illumination, phase effects, and planetary size and albedo. 

We demonstrate our method of estimation by applying it to our understanding of the \textit{TPF-C} as of October 2004. That understanding comprises NASA's working design and the draft science requirements document of the \textit{TPF} science working group. Because neither of these sources is available in published form, we present our understanding as hypotheses adopted to illustrate our method. 

In our understanding, the \textit{TPF-C} working design as of October 2004 is a space-based, diffraction-limited telescope with an unobscured, 8-by-3.5 meter, elliptical, primary mirror, followed by a starlight suppression system and a detector that critically samples the point-spread function at visible wavelengths. The detection zone is an elliptical annulus that extends closest to the star at the two position angles marked by the major axis of the aperture projected onto the target star. At these position angles, which are 180$^{\circ}$ apart, the minimum angular separation of a companion source is the ``inner working angle'' (\textit{IWA}), which is 0.057 arcseconds 
($4\lambda/D$ for $\lambda=550$~nm and $D=800$~cm), and which we assume is valid over a 20\% optical bandwidth 
($\Delta\lambda=110$~nm). The corresponding outer working angle (\textit{OWA}) of the detection zone is 0.684 arcsec. We assume that obtaining and combining observations at three roll angles ($n_\mathrm{r}=3$) of the instrument separated by 60$^{\circ}$ effectively circularizes the inner edge of the detection zone at one epoch. Within the detection zone, assuming sufficiently long total integration times to gather adequate photons, we assume the working design is capable of detecting with \textit{SNR}$~=10$ any point source with a delta magnitude relative to the star less than $\Delta mag_0=25$. This ``limiting delta magnitude'' defines the minimum robustly detected planet or ``limiting planet.'' We assume that $\Delta mag_0$ is controlled by systematic errors, which means that fainter sources are not robustly detectable for any integration time, while all brighter sources are robustly detectable given adequate exposure time. For \textit{SNR} calculations, we assume that the surface brightness of starlight in the detection zone is suppressed to a uniform contrast level $\zeta=5 \times10^{-11}$ with respect to the theoretical Airy peak of the stellar image and that the intensity of extrasolar zodiacal light is three times the solar-system value.

From the draft science requirements document dated March 11, 2004, which is the latest available version in October 2004, we understand that a key planetary population of interest is approximately represented by spheres of Earth's radius with Earth's geometric albedo and exhibiting the Lambert phase function. \textit{TPF-C} target stars should be main-sequence F, G, and K~stars. The possible planets revolve around the target stars on randomly oriented Keplerian orbits that should be drawn from uniform probability distributions in semi-major axis ($a$) over the range 0.7 to 1.5~AU times the square root of the stellar luminosity ($L$) in solar units and in eccentricity ($\varepsilon$) over the range 0 to 0.35, which we call ``habitable orbits."  Assuming every star has exactly one planet of interest, we understand that the \textit{TPF} mission is expected to find 30 or 150 of them after searching 35 stars (minimum mission) or 165 stars (full mission), respectively.

Our demonstrative observing program (1)~ranks target stars according to the expected rate at which planets will be discovered, 
(2)~observes each star only at one, initial epoch, (3)~exposes long enough to reach the limiting planet, and (4)~stops observing new stars after a set total exposure time has elapsed.

Using these simple models of the detection zone, science guidelines, and observing program, we proceed to develop and demonstrate our method of estimating the number of planets discovered. In Section~2, we determine the completeness function for Earth-like planets on habitable orbits as a function of $\Delta mag_0$, $a_0=\textit{IWA}~d$, and {$L$}, where $d$ is the distance to the star in parsecs. In Section~3, we introduce algorithms for calculating the \textit{SNR} of a detection from the exposure time. In Section~4, we define the input pool of possible target stars. In Section~5, we use the demonstrative observing program to compute the expect number of planets found by the working design in one year of exposure time. In Section~6, we illustrate the usefulness of our method in instrument design by performing an optimization of $\Delta mag_0$ for both round and elliptical 8-meter apertures. In Section~7, we use a simple model of an optimized coronagraph to interpret $\Delta mag_0$ in terms of ``wavefront stability." In Section~8, we discuss options for increasing the fidelity and versatility of the models on which our method is based. In Section~9, we discuss options for modeling and verifying the DRM. 

\section{A FUNCTIONAL FORM FOR PHOTOMETRIC AND OBSCURATIONAL COMPLETENESS}

Our simple photometric criterion for detectability in reflected starlight is
\begin{equation}
\Delta mag < \Delta mag_0~~, 
\end{equation}
where $\Delta mag$ is the delta magnitude of any planet. 

In general, 
\begin{equation}
\Delta mag  =  -2.5~\log \frac{F_\mathrm{p}}{F_\mathrm{s}} = -2.5~\log \left( p~\Phi (\beta) \left(\frac{R}{r}\right)^2\right)~~,
\end{equation}
where $F_\mathrm{p}$ and $F_\mathrm{s}$ are the spectral fluxes of the planet and star integrated over $\Delta\lambda$ at $\lambda,~r$ is the distance between the planet and the star, $\beta$ is the phase angle (i.e., the planetocentric angle between the star and the observer), $\Phi(\beta)$ is the planetary phase function, $R$ is the radius of the planet, and $p$ is the geometric albedo of the planet (at $\lambda = 550$~nm, nominally $p_\mathrm{E}\simeq0.33$ for Earth-like and $p_\mathrm{J}\simeq0.50$ for Jupiter-like planets, which we carry in this section out of general interest). Assuming the Lambert phase function,
\begin{equation}
\Phi_\mathrm{L}(\beta) \equiv \frac{\sin \beta + (\pi - \beta)~\cos \beta}{\pi }~~,
\end{equation}
we find for Earth-like and Jupiter-like planets observed at $\lambda = 550$~nm, 
\begin{equation}
\Delta mag_\mathrm{E} = 23.05 - 2.5~\log \left(\frac{\sin \beta + (\pi - \beta)\cos \beta}{\pi }\right) + 5~\log r
\end{equation}
and
\begin{equation}
\Delta mag_\mathrm{J} = 17.36 - 2.5~\log \left(\frac{\sin \beta + (\pi - \beta )\cos \beta}{\pi }\right) + 5~\log r~~,
\end{equation}
where the units of $r$ are AU.

Parallel to our earlier nomenclature for obscurational completeness, we use the name ``ensemble visit photometric and obscurational completeness'' (\textit{EVP\&OC}) for the case of an \textit{ensemble} of planets observed at only one, initial \textit{visit}.

To find \textit{EVP\&OC} as a function of $a_0$ and $\Delta mag_0$, we used the protocols described in Brown (2004a) to prepare a sample of 100 million Earth-like planets on habitable orbits around a star with $L=1$. We drew the orbits from probability distributions uniform in semimajor axis over the range $0.7\leq a \leq 1.5$~AU and in eccentricity over the range $0\leq\varepsilon\leq 0.35$. We randomized the orientations of the orbits and the mean anomalies of the planets. We determined $s$ and $\Delta mag_\mathrm{E}$ for each planetary position. We determined the probability density after sorting the $(s,\Delta mag_\mathrm{E})$ pairs into a $1000\times1000$ grid over the ranges $0\leq s \leq 2.025$~AU and $20\leq\Delta mag_\mathrm{E}\leq40$. Figure~1 shows the resulting distribution of probability density, plotted against the inner scales.  For stars with arbitrary luminosity, the values on the inner abscissa should be read as $s/\sqrt{L}$ or the abscissa values should be multiplied by $\sqrt{L}$, and the ordinate should be read as $\Delta mag -2.5~\log L$ or the ordinate values should be additively increased by $2.5~\log L$. The outer scales show the corresponding distribution for Jupiter-like planets on jovian orbits: the outer abscissa is 7~times the abscissa for habitable orbits with $L=1$, and the outer ordinate is offset upwards by (17.36--23.05)~$-5~\log 7 =1.468$ with respect to the ordinate for habitable orbits, based on equations~(5) and~(6).

The distribution of probability density versus $s$ and $\Delta mag$ has the appearance of a bird in flight. The ``tail" consists of planets located beyond the plane of the sky through the star with separations less than the minimum planet-star distance.  For habitable orbits and $L=1$, this minimum distance is $a_\mathrm{min}~(1-\varepsilon_\mathrm{max})=0.455$~AU, where $a_\mathrm{min}=0.7$~AU is the minimum semimajor axis, and $\varepsilon_\mathrm{max}=0.35$ is the maximum eccentricity. The phase angles of these planets are limited to the range $0\leq\beta  \leq\pi/2$ (fully to half illuminated). The `wing' consists of planets located on the near side of the plane of the sky through the star, which are viewed at phases in the range $\pi/2\leq\beta\leq\pi$. For these planets, there is no lower limit to the brightness, which results in the upward sweep towards smaller separations as the dark hemisphere of the planet is increasingly presented to the observer. The gap between wing and tail is due to the planet-free zone around the star, within 0.455~AU for this planetary population. 

As explained in Brown (2004b), the orange curve is the maximum planet brightness versus separation, which for Lambertian spheres occurs at $\beta=63.3^{\circ}$. From equation~(5) or (6), using $r=s/\sin\beta$, the equation of the orange curve is 
\begin{equation}
\Delta mag_\mathrm{E,~min} = 23.89 + 5~\log s
\end{equation}
for the inner scales, and
\begin{equation}
\Delta mag_\mathrm{J,~min} = 18.20 + 5~\log s
\end{equation}
for the outer scales. Shown by the blue curve, along any line of sight, a planet at dichotomy is 0.4 magnitudes fainter than a planet at brightest phase.

\textit{EVP\&OC} is the integral of the probability density over the ranges of $s$ and $\Delta mag$ in equations~(1) and (2). (We ignore obscuration for $s>\textit{OWA}$ in our analysis of \textit{TPF-C} observing habitable orbits, because it is a small effect relevant only to Alpha Cen~A and~B.) Figure~2 shows our results for \textit{EVP\&OC}, obtained by appropriately summing the probability density on the $1000\times1000$ grid, over the ranges $0\leq a_0\leq2.025$~AU and $20\leq\Delta mag_0\leq40$, but displaying only the range $20\leq\Delta mag_0\leq30$. 

The curves in Figure~2 show the value of $\Delta mag_0$ versus $a_0$ at which \textit{EVP\&OC} is the indicated fraction of its asymptotic value as $\Delta mag_0\rightarrow\infty$. 

We constructed a bi-linear interpolation function for the Earth-like, habitable-orbit case, \textit{fEVP\&OC}$(a_0,~\Delta mag_0)$ for $L=1$. This function produces interpolated values of \textit{EVP\&OC}$_\mathrm{E}$ for $L=1$, in the range of the $1000\times1000$ grid of the Monte Carlo computations. It enables us to estimate \textit{EVP\&OC} for any combination of star and instrument and for either choice of planetary population. For Earth-like planets on habitable orbits:
\begin{equation}
\textit{EVP\&OC}_\mathrm{E} = \textit{fEVP\&OC} \left( \frac{a_0}{\sqrt{L}},~\Delta mag_0 -2.5~\log L \right)~~,
\end{equation}
and for Jupiter-like planets on jovian orbits:
\begin{equation}
\textit{EVP\&OC}_\mathrm{J} = \textit{fEVP\&OC}\left( \frac{a_0}{7},~\Delta mag_0 + 1.468 \right)~~.
\end{equation}

\section{ALGORITHMS TO COMPUTE SIGNAL-TO-NOISE RATIO (\emph{SNR})}

We assume that the observing protocol involves subtracting pairs of images of equal exposure time ($\tau/2$), with any possible planet located at intentionally different positions in the two images, but with no intentional change in the pattern of scattered starlight (speckles). This could be accomplished by a small roll of the telescope around the line of sight between the two exposures.

The photon-statistical \textit{SNR} is:
\begin{equation}
SNR = \frac{C_\mathrm{p}}{\sqrt{C_\mathrm{p} + 2C_\mathrm{b}}}~~,
\end{equation}
where $C_\mathrm{p}$ and $C_\mathrm{b}$ are the numbers of detected planetary and background photons, respectively. For the effective wavelength of the $V$~passband, $\lambda =550$~nm, and fractional bandwidth $\Delta\lambda/\lambda\ll1$, we can estimate $C_\mathrm{p}$:
\begin{equation}
C_\mathrm{p} = \mathcal{F}_0~10^{-\frac{V_\mathrm{s} + \Delta mag}{2.5}}~ \frac{1}{y}~ \frac{\pi}{4}~ D^2~ \varrho^m~ \epsilon~\eta~\Delta\lambda~\tau~~, 
\end{equation}
where $\mathcal{F}_0$ is the $V$-band specific flux for zero magnitude (nominally, 9500 photons cm$^{-2}$ nm$^{-1}$ sec$^{-1}$), 
$V_s$ is the $V$-band apparent magnitude of the star, $y>1$ is the aspect ratio and $D$ is the major axis of the elliptical entrance pupil, $\varrho$ is the reflectivity of the mirrors, $m$ is the number of reflections, $\epsilon$ is the quantum efficiency of the detector, 
$\eta$ is the areal fraction of the clear portion of a possible Lyot-type pupil-plane mask, and $\tau$ is the total exposure time for the two images that are to be differenced.  The two images are possibly obtained by adding multiple shorter exposures. (The calculations show that we are background limited, not read-noise limited.)

To estimate $C_\mathrm{b}$, we must consider contributions from several sources. The background starlight (speckles) contributes:
\begin{equation}
C_\mathrm{b,s} = \mathcal{F}_0~10^{-\frac{V_\mathrm{s}}{2.5}}~\zeta~PSF_\mathrm{Airy~peak}~n_\mathrm{x}~\Omega_\mathrm{x}~\frac{1}{y}~\frac{\pi }{4}~D^2~\varrho^m~\epsilon~\eta~\Delta\lambda~\tau~~, 
\end{equation}
where 
\begin{equation}
PSF_\mathrm{Airy~peak} = \frac{1}{y}~\frac{\pi~D^2}{4\lambda^2}
\end{equation}
is the theoretical Airy peak of the stellar point-spread function, $\zeta$ is defined in Section~1, $n_\mathrm{x}\equiv1/\Psi=1/(0.07\eta)$ is the number of ``noise pixels,'' $\Psi$ is the sharpness, 0.07 is the sharpness for a critically sampled diffraction-limited image from a perfect circular aperture (Burrows 2003; Brown et~al.\ 2003), and $\Omega_\mathrm{x}$ is the solid angle of a detector pixel in steradians (nominally $y(\lambda /2D)^2$ for critical sampling). Zodiacal light contributes:
\begin{equation}
C_\mathrm{b,zl} = \mathcal{F}_0~10^{-\frac{mag\Omega_\mathrm{zl}}{2.5}}~(1+\mu)n_\mathrm{x}~\frac{\Omega_\mathrm{x}}{(4.848\times10^{-6})^2}~\frac{1}{y}~\frac{\pi }{4}~D^2~\varrho^m~\epsilon~\eta~\Delta\lambda~\tau~~,
\end{equation}
where $mag\Omega_\mathrm{zl}$ is the $V$-band intensity of the zodiacal light in magnitudes per square arcsecond (nominally, 
$mag\Omega_\mathrm{zl}=23$), $\mu$ is the brightness of the (unknown) native, extrasolar zodiacal light in units of the solar-system zodiacal light, and $4.848\times10^{-6}$ is the number of radians in an arcsecond. In the calculations of \textit{SNR} in this paper, we assume $\mu=3$.  The dark count contribution is 
\begin{equation}
C_\mathrm{b,d}=\xi~n_\mathrm{x}~\tau~~, 
\end{equation}
where $\xi$ is the dark count rate. The read-noise contribution is 
\begin{equation}
C_\mathrm{b,r} = 2R^2~ n_\mathrm{x}~~, 
\end{equation}
where $R$ is the read noise. (If multiple readouts are performed in the process of obtaining the two images to be differenced, then the multiplier is the total number of readouts. Here, for simplicity, we assume only two readouts, one for each image.  The read noise should be negligible in any case.)

Given values for the astronomical, instrumental, and observational parameters, we can compute $C_\mathrm{b}$. (Table~1 lists the parameters of the \textit{TPF-C} working design relevant to the \textit{SNR} calculation, including reasonable values of the parameters not already discussed.) Then, we can use equation~(11) to compute \textit{SNR} given $\tau$---or vice versa, for the inverse calculation, to find the necessary $\tau$ to achieve a desired value of \textit{SNR}. 

\section{INPUT POOL OF POSSIBLE TARGET STARS}

The \textit{Hipparcos} star catalogue contains a total of 2350 stars within 30~pc, of which 2276 have adequate photometric information  to assign both a $B_\mathrm{s}$--$V_\mathrm{s}$ color and a value of $L$.  Of these, 1408 are main-sequence stars with $B_s$--$V_s > 0.3$ and no stellar companion closer than 10~arcsec according to the Washington Double Star and \textit{Hipparcos} catalogues (Turnbull 2004). Figures~3 to 7 show scatter diagrams for these stars for cases discussed in Sections~5 and~6. The gray points represent the 957 stars with completely obscured habitable orbits: the maximum apparent separation of a planet for these stars is less than $a_0$:
\begin{equation}
s_\mathrm{max} =(a_\mathrm{max}=1.5~\mathrm{AU})\left(1+(\varepsilon_\mathrm{max}=0.35)\right) 
= 2.025~\mathrm{AU} <\frac{a_0 = d (IWA=0.057~\mathrm{arcsec})}{\sqrt{L}}~~.
\end{equation}
The black and colored dots represent the 451 stars for which at least some habitable orbits are unobscured, and they comprise our input target pool.

\section{DEMONSTRATIVE OBSERVING PROGRAM FOR THE WORKING DESIGN OF \emph{TPF-C}}

We represent a \textit{TPF-C} observing program on extrasolar planets by (1)~a pool of possible target stars, (2)~protocols defining the observations (e.g., instrumental settings and exposure times), (3)~rules for selecting the star to observe next, (4)~algorithms for computing the full time cost of an observation, and (5)~a total budget of time for the program. 

In our demonstrative observing program, (1)~the target pool comprises the 451 stars of Section~4; (2)~the observing protocol is to expose for time $\tau$/2 
in each of 2~$n_\mathrm{r}$ exposures, where $\tau$ is the exposure time to achieve $SNR=10$ on the limiting planet ($\Delta mag_0$, with pairs of images at each of $n_\mathrm{r}$ roll angles 180$^{\circ}/n_\mathrm{r}$ apart, with a small change in roll angle between the images of a pair, for subtracting the background scattered starlight; (3)~the rule for star-selection is to choose the unobserved star offering the highest discovery rate ($<$$n_\mathrm{p}$$>\!\!/n_\mathrm{r}\tau$); and (4)~the costing algorithm counts only the total exposure time for a star ($n_\mathrm{r}\tau$), with no overheads.

In this section, we show the detailed results of the demonstrative observing program for ``case~A,'' which is the \textit{TPF-C} working design ($y=8/3.5$, $n_\mathrm{r}=3$, $\Delta mag_0=25$) and a time budget of 12~months (8766~hr). In Section~6, we explore the trade space of $\Delta mag_0$ and ($y$, $n_\mathrm{r}$) for time budgets of 3~months, 6~months, and 12~months, and report ``case~B'' ($y=1$, $n_\mathrm{r}=1$, $\Delta mag_0=25$), ``case~C'' ($y=8/3.5$, $n_\mathrm{r}=3$, $\Delta mag_0=25.45$, ``optimized''), and ``case~D" ($y=1$, $n_\mathrm{r}=1$, $\Delta mag_0=26.1$, ``optimized'').

We compute $<$$n_\mathrm{p}$$>\!\!/(n_\mathrm{r}\tau$) for each star in the pool to prioritize the stars into a rank-ordered target list. We find the denominator ($n_\mathrm{r}\tau$) by solving for the implicit $\tau$ in equation~(11) with $SNR=10$.  We find the numerator ($<$$n_\mathrm{p}$$>$) from the value of the function \textit{EVP\&OC}:
\begin{equation}
<\!\!n_\mathrm{p}\!\!> \pm~\Delta\!<\!\!n_\mathrm{p}\!\!>\, =\textit{fEVP\&OC}_\mathrm{E} \pm \sqrt{\textit{fEVP\&OC}_\mathrm{E} 
(1\!-\!\textit{fEVP\&OC}_\mathrm{E})}~~.
\end{equation}

Figure 8A shows a scatter plot of the discovery rate for the input pool of target stars.

We compute the total time budget:
\begin{equation}
T\equiv\sum_{i=1}^{n_s} n_\mathrm{r}\tau(i)\!=\!12~\mathrm{months}~~,
\end{equation}
where $n_\mathrm{s}$ stars are indexed by $i$ in rank order of their priority. We compute the expectation value and standard deviation of the grand total of found planets ($\Sigma n_\mathrm{p}$):
\begin{equation}
<\!\!\Sigma n_\mathrm{p}\!\!>\!\pm~\Delta\!\!<\!\!\Sigma n_\mathrm{p}\!\!>\,=  \sum_{i=1}^{n_\mathrm{s}} \textit{fEVP\&OC}_\mathrm{E}(i)~ \pm \sqrt{\sum_{i=1}^{n_\mathrm{s}} \textit{fEVP\&OC}_\mathrm{E}(i) (1\!-\!\textit{fEVP\&OC}_\mathrm{E}(i))}~~. 
\end{equation}

Table 2 shows the detailed results of the demonstrative observing program for case~A. We expect $26.9\pm4.2$ planets to be discovered in a year of exposure time, observing 117~stars and assuming each star has exactly one Earth-like planet on a habitable orbit.

\section{AN OPTIMIZATION OF $\mathbf{\Delta}$\emph{mag}$\mathbf{_0}$ FOR ROUND AND ELLIPTICAL 8-METER APERTURES}

The purpose of this section is to illustrate the usefulness of our method of estimating the yield of search programs for instrument design. We use variations of the demonstrative observing program to explore the optimization of $\Delta mag_0$, perhaps the most critical specification of the instrument, for various values of grand total exposure time. Here, we consider both round and elliptical 8-meter apertures. In Section~7, we use a simple model of an optimized coronagraph to provide one interpretation $\Delta mag_0$, in terms of ``wavefront stability.''

Except for the following changes, we use here the same demonstrative observing program as in Section~5. We allow $\Delta mag_0$ to be a free parameter, allow ($y$, $n_\mathrm{r})=(1,1)$ or (8/3.5, 3) for the round and elliptical cases, respectively, and allow $T=3$~months, 6~months, or 1~year. Figure~9 shows the variation of $<$$\Sigma n_\mathrm{p}$$>$ with $\Delta mag_0$ for these cases.   Table~3 tabulates the results for both the optimal values of $\Delta mag_0$ and $\Delta mag_0=25$. 

The peaks in Figure 9 occur where the gains from searching deeper exactly balance the losses from dropping stars to find the time to search more deeply. The peaks occur at higher values of $\Delta mag_0$ for higher values of $T$ because when those gains and losses are balanced, an increase of alloted time is better spent searching more deeply on higher priority stars than searching a greater number of lower priority stars.

Figures 3--8 show how the demographics and completenesses of the observed stars vary with aperture shape and limiting sensitivity for the demonstrative observing program. When $\Delta mag_0$ is reduced from the value that maximizes the number of planets found, the completeness drops, more low-yield stars are searched, and the discovery rate increases for high-completeness stars and decreases for low-completeness stars. (Another negative consequence of lower $\Delta mag_0$ is that any found planet will be observable over a smaller portion of its orbit.) Compared to the optimized round aperture, the optimized elliptical aperture offers lower completeness, observes fewer stars, and finds fewer the planets.

\section{AN INTERPRETATION OF $\mathbf{\Delta}$\emph{mag}$\mathbf{_0}$ IN TERMS OF WAVEFRONT STABILITY}

The stability of the speckles that constitute the scattered starlight in the detection zone may set the systematic limit to the sensitivity, which
is expressed by $\Delta mag_0$ (Brown 1988; Brown \& Burrows 1990). The speckles are due to residual wavefront errors. Assuming speckle stability \textit{is} the limiting factor, we can use an engineering model of the instrument to translate values of $\Delta mag_0$ into specifications of wavefront stability relevant to the design of \textit{TPF-C}. We present one possible definition of ``speckle stability'' and one possible translation of $\Delta mag_0$ using the simplified model of an optimized coronagraph in Brown et~al.\ (2003).

In our model, wavefront errors are corrected by a deformable mirror with $2N\times2N$ actuators, with rows and columns aligned with
the major and minor axes of the aperture. The deformable mirror has authority over spatial frequencies between 1 and $N$~cycles across the major and minor axes. Assume residual wavefront errors are isotropic. Assume they are approximately uncorrelated, which means the power spectral density (PSD) is approximately flat, as was observed in the laboratory by Trauger et~al.\ (2003) for a deformable mirror technology under study for \textit{TPF-C}. Then, from Elson (1984) and Elson et~al.\ (1983), as developed further in Brown \& Burrows (1990) and Brown et~al.\ (2003), the point-spread function of scattered starlight in the detection zone due to wavefront errors is approximately uniform with the value:
\begin{equation}
\mathit{PSF}_\mathrm{PSD}=\frac{4\pi~\sigma^2~D^2}{\lambda^4~N^2} \equiv
\zeta \frac{1}{y} \frac{\pi~D^2}{4\lambda^2}~~, 
\end{equation}
where $\sigma^2$ is the integral of the PSD over the controlled range of spatial frequencies for the major axis, and the triple equality is
the definition of the contrast level $\zeta$, from which we can derive:
\begin{equation}
\zeta=\frac{16~y~\sigma^2}{N^2~\lambda^2}~~,
\end{equation}

It is our understanding that the working design for \textit{TPF-C} in October 2004 has $N=48$, so the mean-square wavefront error implied by the contrast level cited in Section~1, $\zeta=5\times10^{-11}$, is $\sigma^2=(0.0308)^2$~nm$^2$.

From equations (12) and (13), the delta magnitude of the planet that produces the same counts as the background produces in 
$n_\mathrm{x}$ noise pixels is 
\begin{equation}
\Delta mag_\mathrm{sp}=-2.5 \log \frac{\pi~\sigma^2~y}{\lambda^2~N^2~0.07\eta} 
=-2.5 \log \frac{\pi~\zeta}{16~0.07\eta}~~,
\end{equation}
and $\Delta mag_\mathrm{sp}=23.88$ for the working design.

We parameterize the stability of speckles by the upper limit to an additional, systematic wavefront error, with the mean-square value $\delta\sigma^2$, that applies to only one of the two images in a pair to be differenced. Assuming the wavefront errors described by $\sigma^2$ and $\delta\sigma^2$ are independent, their squares add in first equation~(22) when determining $PSF_\mathrm{PSD}$ for the image that is differentially affected by $\delta\sigma^2$. Therefore, the delta magnitude of a ``systematic speckle"---one that does \textit{not} cancel in the observing protocol and is therefore a source of systematic error and confusion---is:
\begin{equation}
\Delta mag_\mathrm{ssp} = \Delta mag_\mathrm{sp} - 2.5~\log \frac{\delta\sigma^2}{\sigma^2}~~. 
\end{equation}

The delta magnitude of the limiting planet ($\Delta mag_0$) must be a small multiple of the brightness of a systematic speckle (nominally $4\times$ brighter or 1.5~magnitudes, which could be optimized by simulations of the signal recovery process). Therefore, nominally:
\begin{equation}
\Delta mag_0 = -2.5~\log \frac{\pi~\sigma^2~y}{\lambda^2~N^2~0.07\eta}  - 2.5~\log \frac{\delta\sigma^2}{\sigma^2} - 1.5~~,
\end{equation}
In Table 3, we have used equation~(26) to determine the constraint on $\delta\sigma^2$ from $\Delta mag_0$ for the cases treated in Sections~5 and~6. 

\section{OPTIONS FOR INCREASING THE FIDELITY AND VERSATILITY OF THE MODELS}

To this point, we have sailed before the wind, marshaling ``understandings,'' ad hoc models, and a working design to demonstrate a simple yet entire estimate of the productivity of a search program for extrasolar planets. We have succeeded in establishing an initial relationship between the size, quality, and stability of the instrument's optics and its ability to meet mission science requirements. We have forged a preliminary tool, and in this section, we discuss how to increase its usefulness to the \textit{TPF-C} project by improving its component parts. And in the next section, we discuss extensions of the method for mission-level verifications.

\textit{Planetary population of interest}. The draft \textit{TPF} science requirements document states, ``\textit{TPF} must be able to detect
terrestrial planets different from our own, down to a minimum terrestrial planet defined as having 1/2 Earth surface area...,'' but the document provides no further guidance, for example about the ``maximum'' terrestrial planet. This is an issue for our method, because the starting point for completeness calculations is a full description of the planets being sought in terms of probability distributions. In the case of size, the need could be satisfied by a minimum and maximum planetary radius and an assumed power-law distribution, such as a uniform distribution in either radius or surface area. Possibly the planetary albedo should be represented by a probability distribution, also. In any event, for purposes of optimizing the instrument and verifying the mission, the planets of interest must be fully represented by Monte Carlo ensembles drawn from relevant planetary properties.

\textit{Selection of the pool of target stars}. We can expect vigorous debate on the selection criteria for target stars to search for planets,
and we can hope for ample theoretical and observational research to inform those choices. In addition to evidence of stellar singularity, main-sequence occupancy, and greater than billion-year life expectancy, which motivated the criteria in Section~4, we might also consider the presence of giant planets found by radial-velocity or astrometric techniques, circumstellar features, such as ionizing radiation or circumstellar dust, and other characteristics.

\textit{The detection zone and SNR calculations}. A detection zone of a real instrument will not have a sharp inner edge and uniform sensitivity, nor will the line between ``detectable'' and ``undetectable'' be as sharply drawn as in this paper, with an abrupt transition between a regime of photon statistics and a regime blocked by systematic effects. We must draw from the instrumental design valid algorithms for computing \textit{SNR} from the planetary flux, the focal plane position of the planet, and the level of extrasolar zodiacal light. A more sophisticated treatment could vary the solar-system zodiacal light with ecliptic latitude, but this refinement would be small compared with the current uncertainty in $\mu$.  We should also evaluate critically our current criterion that $SNR=10$ is the sharp line between ``detected'' and ``undetected.'' The likely complexity of these considerations will rule out the simplification of a functional description of completeness and will require keeping track of individual possible planets in a Monte Carlo sample by means of a \textit{computer orrery}.

\textit{Multiple observing epochs}. Another need for a computer orrery is to support multiple observations of the same star at separated times. Because only a subset of the possible planets are detectable at any epoch, multiple observations are needed to accumulate completeness. As time proceeds, the planets that have been either obscured or too faint to detect a previous epochs will move to new positions and change in brightness, possibly becoming detectable. Each possible planet in a Monte Carlo ensemble  moves according to Keplerian theory on the basis of its unique orbital elements, and the computer orrery must keep track of them all, tagging ones that are detected and reporting completeness as the accumulating detected fraction of the whole sample. Brown (2004a) reported results of such multiple-epoch completeness calculations for obscurational completeness alone. Those results show that an optimal delay exists for scheduling each observing epoch after the first, the time delay that maximizes the increase in completeness. The optimal delay depends on the stellar mass, the projected detection zone, and the current ensemble of as-yet-undetected possible planets.

\textit{Full accounting of time costs}. A mission must be planned in clock or calendar time, not exposure time. From the observatory design, we need algorithms for computing the overhead times for slewing, setup, and housekeeping, which must be added to each exposure time to obtain the realistic estimates of time costs.

\textit{Planning and scheduling algorithm}.  We need a formalism to determine the next observation after the current one. This formalism must comprise both a planning and scheduling algorithm, which suggests candidates for the next observation based on technical factors, and a human authority that resolves conflicts when technical factors do not make the choice clear. In this paper, we have used a very simple planning and scheduling algorithm, which prioritized stars only for a first searching observation based on predicted discovery rate alone. The natural extension for multiple observing epochs calls for a computer orrery to provide current discovery rate estimates, taking previous observations into account. The planning and scheduling algorithm should also incorporate other types of observations (e.g., planet validation, planet characterization, general astrophysics), constraints and restrictions (e.g., solar avoidance), and the angular speed of the star with respect to the distant background, which facilitates the disambiguation of confusion sources.

\textit{Instrument model for interpreting} $\Delta mag_0$. A improved model of the instrument will take into account the time evolution of the wavefront and its interactions with imperfect masks and optics. The simple model in Section~7 establishes an initial, basic connection
between $\Delta mag_0$ and the size, optical quality, and stability of the instrument.

\section{OPTIONS FOR MODELING AND VERIFYING THE DRM}

Our demonstrative observing program served its purpose, which was to illustrate a new method for estimating the ability of an instrumental design to directly discover extrasolar planets of interest. However, that program is neither an adequate concept for an actual observing program nor a sufficient framework for verifying that a mission design can meet science requirements. It still would not be an adequate framework if the component models achieve satisfactory fidelity and versatility by following the suggestions in Section~8. \textit{The highly contingent nature of an observing program for extrasolar planets demands a new level of simulations for mission verification}. This new level must involve Monte Carlo simulations of the mission as a whole.

Contingency is perhaps the most distinctive feature of an observing program to discover and study extrasolar planets directly. Systems and procedures to manage contingent factors will dominate science operations, and stochastic factors will shape the course of the mission. By examining Monte Carlo samples of the whole mission, we can explore the ranges of mission outcomes for ranges of the unknown factors, and we can hope for confidence in the mission's integrity as a result.

The contingent factors fall into ``strategic'' and ``tactical'' categories. Three strategic factors are (1)~the frequency of occurrence of planets of interest, (2)~the areal density of astronomical confusion sources that must be disambiguated from planets, and (3)~the level of extrasolar zodiacal light around target stars, which for \textit{TPF-C} will dominate the noise for $V_\mathrm{s}\gtrsim6.3$ if the level is only as great as the solar system's zodiacal light. Factors~(2) and~(3) are individual characteristics of each potential target star. 

The tactical contingent factors are the outcomes of previous observations of each star. These results will suggest the type and timing of future observations of that star. If a source is found, it must be validated as a physical companion by a time-critical observing protocol to measure its apparent motion. If a source is validated as a physical companion, it must be further validated as a planet of interest, based on brightness and apparent motion. If it is so verified, then it must be characterized by a further observing protocol, presumably involving spectroscopy, astrometry, and photometry. These validating and characterizing observations for identified sources will demand escalating amounts of time, and for this reason the logic and the criteria for the progressive steps must be well thought out. Interestingly, even null search results are an important contingent factor, because we can use our computer orrery to know exactly which possible planets could not yet have been discovered and to predict how many of them will have, at any future time, revolved in their orbits to have sufficiently improved their brightness and location on the focal plane to now be detectable.

Let us envision that we have a testing environment that comprises adequate models according to Section~8, including a planning and scheduling algorithm to manage the tactical contingent factors and a human authority to resolve scheduling conflicts and select the next observation from technically equivalent options. With this capability in place, we are prepared to explore the range of mission outcomes that follow from assumptions about the strategic contingent factors. These assumptions take the form of probability distributions for 
(1)~the presence of a planet of interest, (2)~the presence of background confusion sources, and (3)~the level of extrasolar zodiacal light. Item~(1) is a Bernoulli random variable with the probability of a positive outcome equal to the assumed occurrence rate of planets of interest. Item~(2) is a Poisson random variable according to the product of the assumed areal density of background sources in the delta magnitude range corresponding to planets of interest and the area of the detection zone. The flux of the background confusion sources could be described, say, by a power-law distribution of delta magnitudes over some range. Item~(3) could be described, say, by a power-law distribution of intensity over a finite range, from a lower limit of a fraction of the solar system's value to an upper limit provided by lack of an observed infrared excess. (If the star is known or can be inferred to have a zodiacal light, that observational value should be used.) 

The first step in a whole-mission simulation---the DRM---is to produce realizations of the strategic contingent factors from their probability distributions. This means generating for each star in the pool (1)~the number of planets, 0 or 1, and if 1, then exactly \textit{which} planet in the computer orrery, (2)~the locations and fluxes of nearby confusion sources, and (3)~the level of extrasolar zodiacal light ($\mu$), The second step, is to compute the entire mission schedule, observation by observation, with the outcome of each observation determined by the instrumental and observational parameters operating on the stellar parameters and the realization. As the mission trial progresses, the next observation is always decided by the planning and scheduling algorithm and surrogate human authority on the basis of all the information at hand, including the results of the previous observation. If the mission has prescribed duration, then the mission outcome comprises the numbers of planets discovered, verified, and characterized after that time has elapsed. If the mission has unlimited time, the simulation will provide results as a function of time. Repeating the two steps many times will yield an empirical probability distribution of mission outcomes, which should be an adequate basis for mission verification.

We can anticipate the qualitatively different characters of the mission depending on the strategic contingent factors. For example, if the occurrence rate of planets is high, then we could expect characterizing observations to dominate the observing program, because they demand much more time than searching observations. If the occurrence rate is low but the confusion background is high, then we could expect validating observations to limit the rate of discoveries. If the occurrence rate and the background confusion are both low, the program will proceed from star to star based on which available star offers the largest increase in completeness. If the exozodiacal light is higher, the exposure times to achieve the same \textit{SNR} will get longer, which will decrease the number of planets studied in a mission of limited duration.

For \textit{TPF-C}, when this testing process instills adequate confidence that one feasible strawman design can achieve the science requirements, the project can move forward from the pre-Phase A to the Phase A study.

\begin{acknowledgements}
We are grateful to Wesley Traub, Stuart Shaklan, Sally Heap, and Donald Lindler for their critical reading of our work on photometric completeness, and to Stuart for ideas on the tolerancing of \textit{TPF}. We acknowledge a long collaboration with Christopher Burrows on the systematic effects of speckles in coronagraphic planet detection. We thank Margaret Turnbull for her thoughtful work and support on target lists. We thank Robert Vanderbei for discussions of statistics. We salute Christian Lallo for his expertise aiding the computations and his craftsmanship rendering the results of this research. JPL contract 1254081 with the Space Telescope Science Institute provided support.
\end{acknowledgements}

\clearpage
\begin{figure}
\epsscale{.7}
\plotone{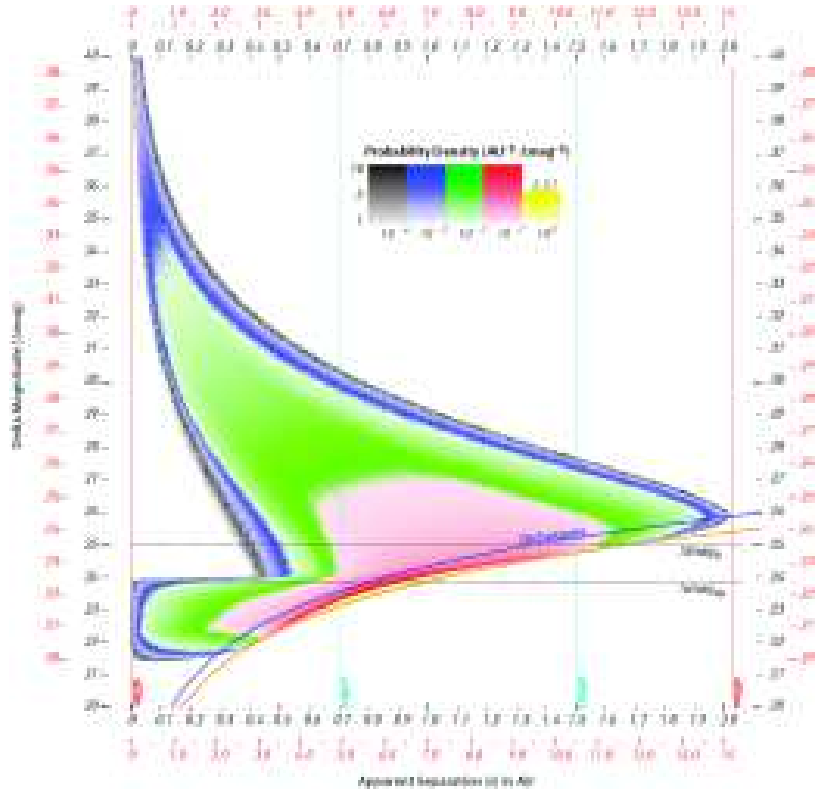}
\caption{The distributions of probability density for two distributions of extrasolar planets versus separation and brightness relative to the star. Based on a Monte Carlo study with 100 million trials. Inner scales: Earth-like planets on habitable orbits around a star with $L=1$. For stars with arbitrary luminosity, the values on the inner abscissa should be read as $s/\sqrt{L}$ or the abscissa values should be multiplied by $\sqrt{L}$, and the ordinate should be read as $\Delta mag -2.5~\log~L$ or the ordinate values should be additively increased by $2.5~\log~L$. Outer scales: Jupiter-like planets on jovian orbits. The orange curve is the maximum brightness versus separation, which for Lambertian spheres occurs at $\beta=63.3^{\circ}$. The blue curve is for planets at dichotomy ($\beta=90^{\circ}$). The lower horizontal line shows the typical delta magnitude of fixed speckles, and the upper horizontal line shows the value of the limiting delta magnitude we assumed for \textit{TPF-C.}}
\end{figure}

\begin{figure}
\epsscale{.7}
\plotone{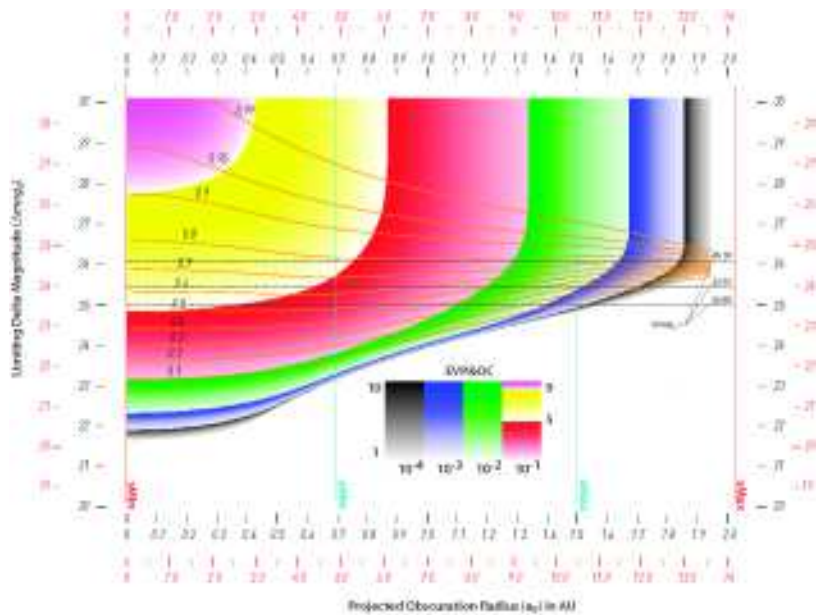}
\caption{\textit{EVP\&OC}, single-visit searching completeness, as a function of projected inner working angle ($a_0$) and limiting delta magnitude ($\Delta mag_0$). Inner scales: Earth-like planets on habitable orbits around a star with $L=1$. For stars of luminosity $L$, the values on the inner abscissa should be read as $a_0/\sqrt{L}$ or the abscissa values should be multiplied by $\sqrt{L}$, and the ordinate should be read as $\Delta mag -2.5~\log~L$ or the ordinate values should be additively increased by 2.5~$\log~L$. Outer scales: Jupiter-like planets on jovian orbits. The orange curves show the values of $\Delta mag_0$ versus $a_0$ at which \textit{EVP\&OC} is the indicated fraction of its asymptotic value as $\Delta mag_0\rightarrow\infty$.  The horizontal dashed lines shows $\Delta mag_0$ for the cases of an 8-meter-class coronagraph summarized in Table~3.}
\end{figure}
\clearpage

\begin{figure}
\epsscale{.75}
\plotone{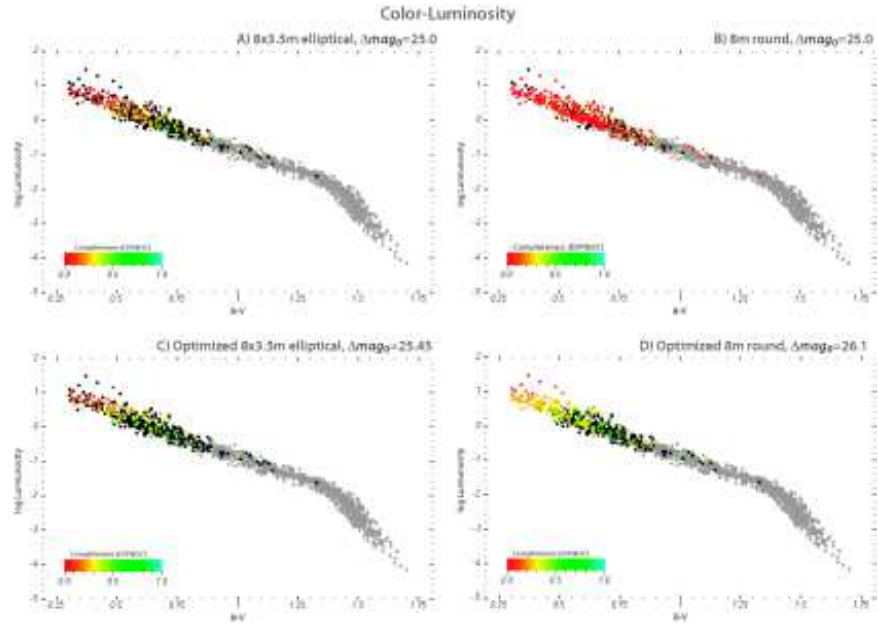}
\caption{Scatter diagrams of color and log luminosity for 1408 are main-sequence stars with $B_s-V_s>0.3$ and no stellar companion closer than 10~arcsec according to the Washington Double Star and \textit{Hipparcos} catalogues (Turnbull 2004). An 8-meter-class coronagraph completely obscures the habitable orbits of 957 stars (gray). The black and colored dots represent the 451 stars for which at least some habitable orbits are unobscured, which comprise the input target list. The four panels represent the cases in the trade-space study of wavefront stability for 8-meter-class coronagraphs with a one-year fixed budget of exposure time used to search for Earth-like planets on habitable orbits.}
\end{figure}

\begin{figure}
\epsscale{.75}
\plotone{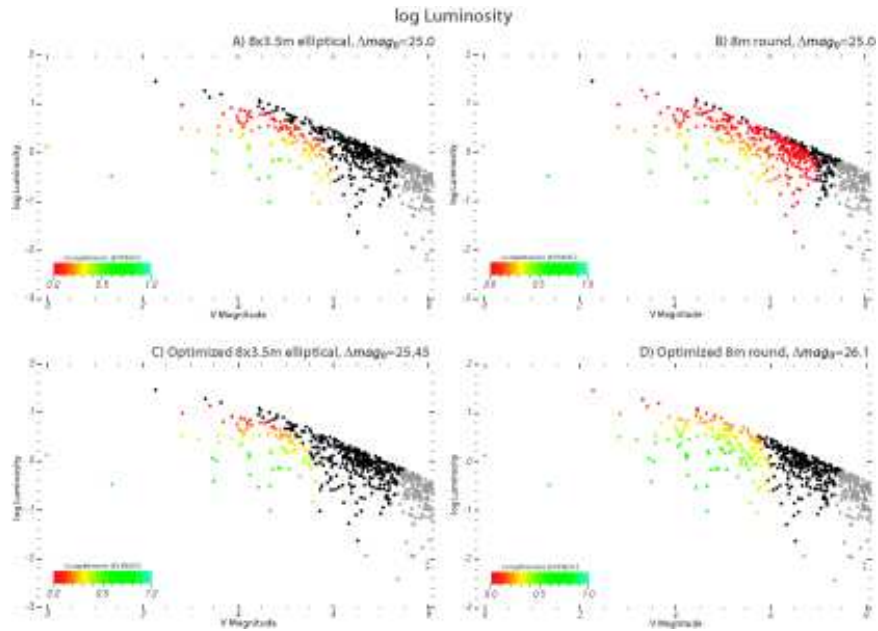}
\caption{Scatter diagrams of log luminosity and magnitude $V_\mathrm{s}$.}
\end{figure}
\clearpage

\begin{figure}
\epsscale{.75}
\plotone{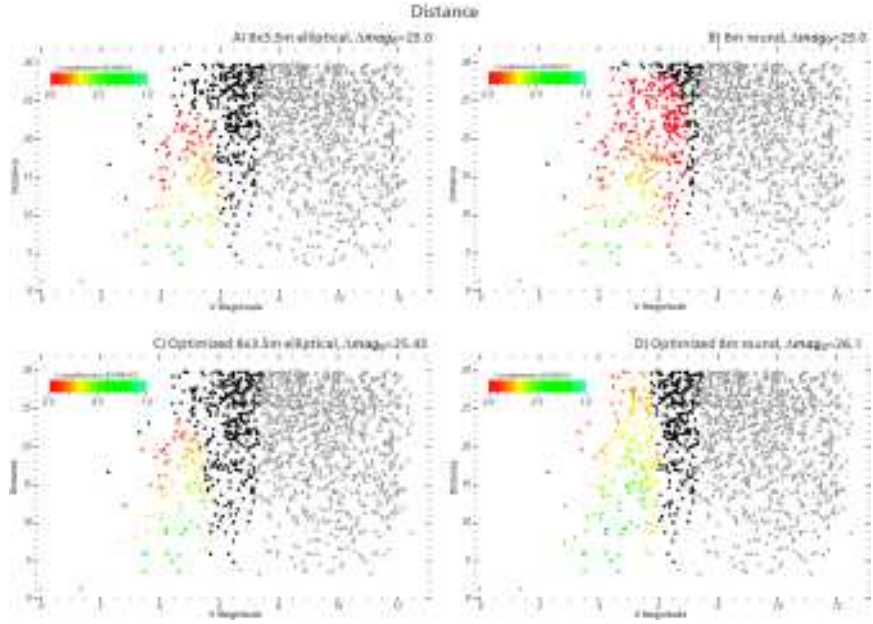}
\caption{Scatter diagrams of distance and magnitude $V_\mathrm{s}$.}
\end{figure}

\begin{figure}
\epsscale{.75}
\plotone{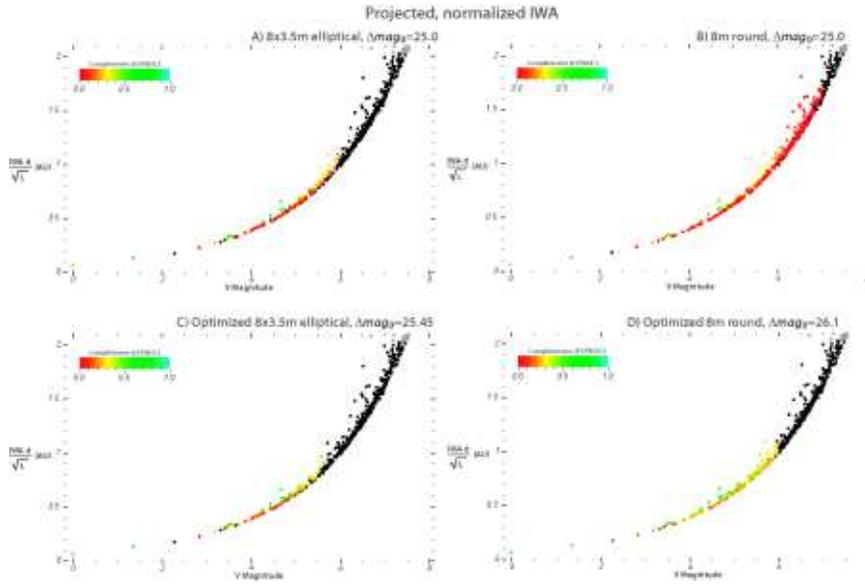}
\caption{Scatter diagrams of projected, normalized \textit{IWA} and magnitude $V_\mathrm{s}$.}
\end{figure}
\clearpage

\begin{figure}
\epsscale{.75}
\plotone{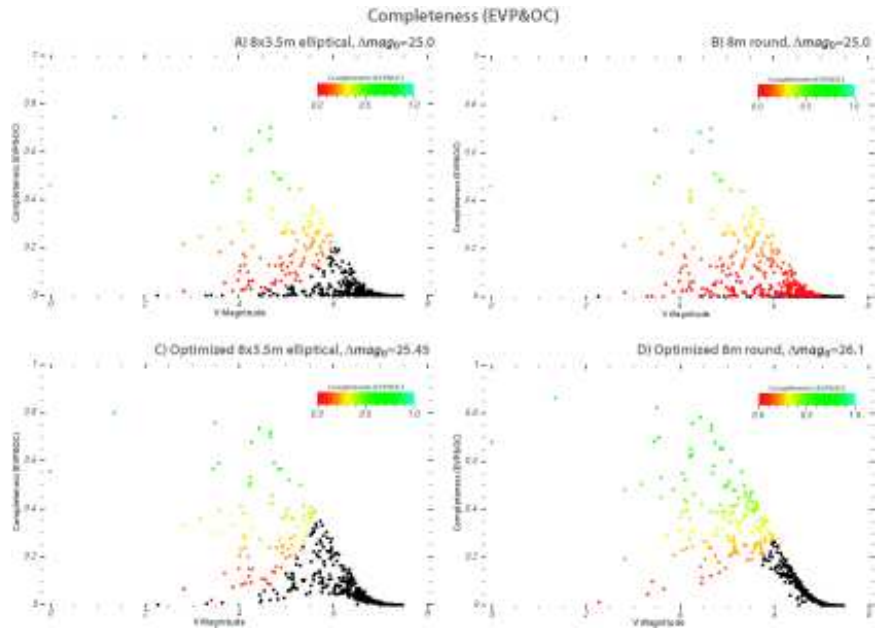}
\caption{Scatter diagrams of completeness (\textit{EVP\&OC}) and magnitude $V_\mathrm{s}$.}
\end{figure}

\begin{figure}
\epsscale{.75}
\plotone{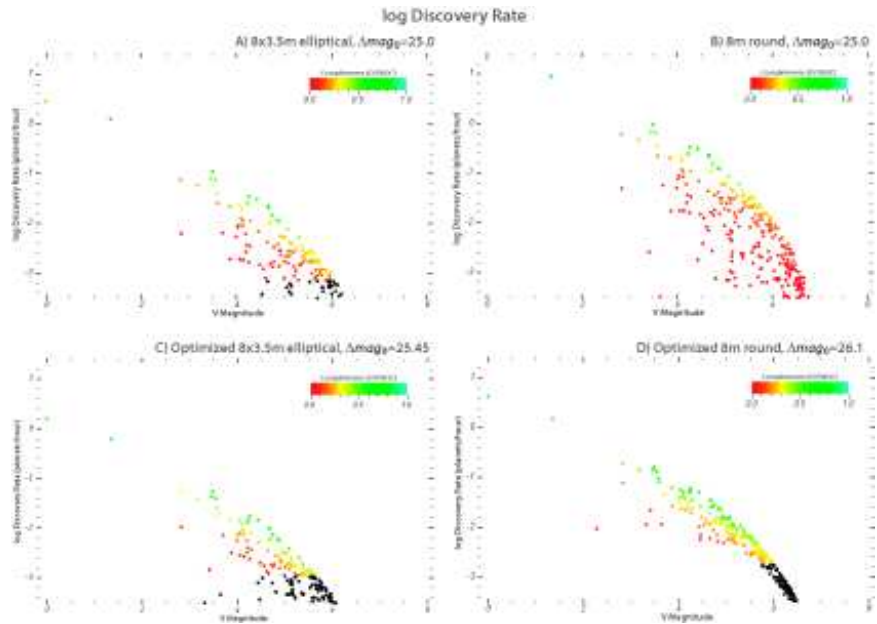}
\caption{Scatter diagrams of log discovery rate and magnitude $V_\mathrm{s}$.}
\end{figure}
\clearpage

\begin{figure}
\epsscale{.75}
\plotone{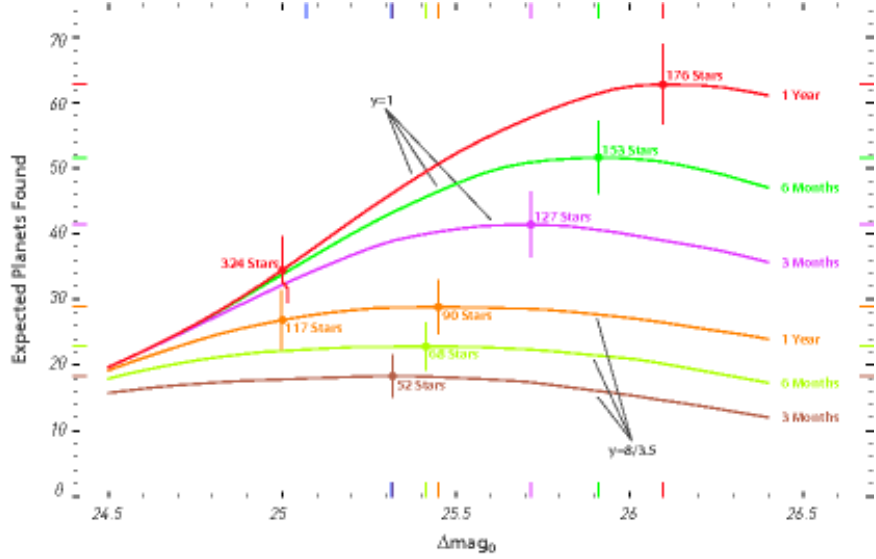}
\caption{Optimization of $\Delta mag_0$. Expectation value of the number of planets found (assuming all stars have one planet drawn from the Earth-like, habitable population) versus the delta magnitude of the limiting planet, for round and elliptical 8-meter coronagraphs observing the input target stars in order of their priority according to discovery rate, with the grand total exposure time limited to $T=3$, 6, and 12~months. The number of stars observed in that time is indicated. The data points are $<$$\Sigma n_\mathrm{p}$$>\pm\Delta<$$\Sigma n_\mathrm{p}$$>$ for the optimal value of $\Delta mag_0$ and $\Delta mag_0=25$. The results are summarized in Table~3.}
\end{figure}

\begin{deluxetable}{llp{4in}}
\tablecaption{Specifications of an 8-meter-class Coronagraph Used in this Paper}
\tabletypesize{\footnotesize}
\tablewidth{0pt}
\tablehead{\colhead{Symbol} &\colhead{Value} &\colhead{Quantity}}
\startdata
$D$ &800 &major axis of aperture in cm\\
$y$ &\phn\phn1 or 8/3.5\tablenotemark{a} &aspect ratio, major axis/minor axis\\
$\zeta$ &\phn\phn$5\times10^{-11}$ &uniform contrast level in detection zone\\
$n_\mathrm{r}$ &\phn\phn1 or 3\tablenotemark{a} &number of roll angles to circularize detection zone\\
$\varrho$ &\phn\phn0.92 &reflectivity of the mirrors\\
$m$ &\phn10 &number of reflections\\
$\epsilon$ &\phn\phn0.8 &quantum efficiency of the detector\\
$\eta$ &\phn\phn0.5 &areal fraction of the clear portion of a possible Lyot-type pupil-plane mask\\
$\Psi$ &\phn\phn0.035 &sharpness, 0.07$\eta$\\
$n_\mathrm{x}$ &\phn28.6 &noise pixels, 1/$\Psi$\\
$\Omega_\mathrm{x}$ &\phn\phn$1.18\times10^{-15}$ &solid angle in steradians of critically sampling pixels at $\lambda=550$~nm\\
$\xi$ &\phn\phn0.001 &dark count rate in sec$^{-1}$ pixel$^{-1}$\\
$R$ &\phn\phn2 &read noise in pixel$^{-1}$\\
$\lambda$ &550 &central wavelength of passband in nm\\
$\Delta\lambda$ &110 &width of passband in nm\\
$\Delta mag_0$ &\phn25\tablenotemark{a} or Table 2 &delta magnitude of the minimum robustly detectable planet or limiting planet\\
\textit{IWA} &\phn\phn0.057 &inner working angle in arcsec, inner limit of detection zone\\
\enddata
\tablenotetext{a}{\textit{TPF-C} working design as of October 2004}
\end{deluxetable}
\clearpage

\LongTables
\begin{deluxetable}{rrlllrcccc}
\tablewidth{0pt}
\tablecaption{}
\tabletypesize{\scriptsize}
\tablehead{
\colhead{Rank} 
&\colhead{HIP} 
&\colhead{$V_\mathrm{s}$} 
&\colhead{$a_0/\sqrt{L}$} 
&\colhead{$<$$n_p$$>$} 
&\colhead{$n_\mathrm{r}\tau$} 
&\colhead{$(<$$n_\mathrm{p}$$>$/$(n_\mathrm{r}\tau))^{-1}$} 
&\colhead{$\Sigma n_\mathrm{r}\tau$} 
&\colhead{$<$$\Sigma n_\mathrm{p}$$>$} 
&\colhead{$\Delta<$$\Sigma n_\mathrm{p}$$>$}\\
&& &\colhead{(AU)} & &\colhead{(hr)} &\colhead{(hr)} &\colhead{(hr)} && }
\startdata
	1. &71683 &\llap{$-$}0.01 &0.066 &0.46 &0.147 &\phn\phn0.319 &\phn\phn0.147 &\phn0.46\phn &0.498 \\
	2. &71681 &1.35 &0.132 &0.744 &0.545 &\phn\phn0.733 &\phn\phn0.692 &\phn1.204 &0.663 \\
	3. &8102 &3.49 &0.331 &0.695 &5.866 &\phn\phn8.436 &\phn\phn6.558 &\phn1.899 &0.807 \\
	4. &3821 &3.453 &0.313 &0.472 &5.596 &\phn11.845 &\phn12.154 &\phn2.372 &0.949 \\
	5. &2021 &2.82 &0.236 &0.214 &2.599 &\phn12.143 &\phn14.753 &\phn2.586 &1.034 \\
	6. &99240 &3.55 &0.343 &0.498 &6.337 &\phn12.722 &\phn21.09\phn &\phn3.084 &1.148 \\
	7. &22449 &3.177 &0.269 &0.245 &3.97\phn &\phn16.208 &\phn25.06\phn &\phn3.329 &1.226 \\
	8. &27072 &3.586 &0.326 &0.282 &6.637 &\phn23.523 &\phn31.696 &\phn3.611 &1.306 \\
	9. &15510 &4.26 &0.469 &0.606 &16.744 &\phn27.625 &\phn48.44\phn &\phn4.217 &1.394 \\
	10. &19849 &4.43 &0.528 &0.686 &21.472 &\phn31.297 &\phn69.912 &\phn4.903 &1.47\phn \\
	11. &1599 &4.23 &0.447 &0.439 &16.035 &\phn36.496 &\phn85.946 &\phn5.342 &1.551 \\
	12. &57757 &3.592 &0.332 &0.182 &6.694 &\phn36.863 &\phn92.641 &\phn5.524 &1.598 \\
	13. &64394 &4.243 &0.451 &0.412 &16.348 &\phn39.651 &108.989 &\phn5.936 &1.672 \\
	14. &105858 &4.219 &0.436 &0.397 &15.773 &\phn39.745 &124.762 &\phn6.333 &1.743 \\
	15. &14632 &4.049 &0.414 &0.307 &12.402 &\phn40.419 &137.163 &\phn6.64\phn &1.803 \\
	16. &78072 &3.85 &0.368 &0.221 &9.428 &\phn42.754 &146.591 &\phn6.86\phn &1.85\phn \\
	17. &108870 &4.667 &0.657 &0.704 &30.674 &\phn43.588 &177.265 &\phn7.564 &1.905 \\
	18. &96100 &4.665 &0.58 &0.65 &30.568 &\phn47.032 &207.833 &\phn8.214 &1.964 \\
	19. &12777 &4.103 &0.415 &0.273 &13.375 &\phn49.068 &221.208 &\phn8.487 &2.014 \\
	20. &64924 &4.74 &0.585 &0.513 &34.312 &\phn66.83\phn &255.52\phn &\phn9.\phn\phn\phn &2.075 \\
	21. &7513 &4.096 &0.417 &0.174 &13.256 &\phn76.308 &268.776 &\phn9.174 &2.109 \\
	22. &15457 &4.845 &0.609 &0.488 &40.379 &\phn82.689 &309.155 &\phn9.662 &2.168 \\
	23. &116771 &4.123 &0.419 &0.164 &13.761 &\phn84.027 &322.916 &\phn9.826 &2.199 \\
	24. &16852 &4.288 &0.458 &0.205 &17.431 &\phn84.951 &340.347 &10.031 &2.236 \\
	25. &57443 &4.89 &0.619 &0.486 &43.336 &\phn89.155 &383.683 &10.517 &2.291 \\
	26. &23693 &4.704 &0.548 &0.359 &32.466 &\phn90.494 &416.149 &10.876 &2.341 \\
	27. &24813 &4.685 &0.558 &0.323 &31.53\phn &\phn97.65\phn &447.679 &11.199 &2.387 \\
	28. &70497 &4.043 &0.404 &0.122 &12.303 &101.022 &459.982 &11.321 &2.409 \\
	29. &102485 &4.129 &0.415 &0.129 &13.872 &107.654 &473.855 &11.449 &2.432 \\
	30. &59199 &4.012 &0.388 &0.102 &11.776 &115.713 &485.63\phn &11.551 &2.451 \\
	31. &29271 &5.075 &0.69 &0.447 &58.149 &129.942 &543.779 &11.999 &2.501 \\
	32. &71284 &4.454 &0.478 &0.157 &22.251 &141.28\phn &566.03\phn &12.156 &2.527 \\
	33. &28103 &3.701 &0.337 &0.053 &7.717 &146.594 &573.748 &12.209 &2.537 \\
	34. &77952 &2.831 &0.225 &0.018 &2.631 &147.687 &576.379 &12.227 &2.541 \\
	35. &112447 &4.193 &0.432 &0.098 &15.21\phn &155.473 &591.589 &12.324 &2.558 \\
	36. &47592 &4.92 &0.608 &0.262 &45.437 &173.634 &637.026 &12.586 &2.595 \\
	37. &50954 &3.986 &0.385 &0.064 &11.367 &176.508 &648.393 &12.651 &2.607 \\
	38. &53721 &5.031 &0.653 &0.303 &54.137 &178.685 &702.531 &12.954 &2.647 \\
	39. &5862 &4.964 &0.626 &0.262 &48.707 &185.848 &751.238 &13.216 &2.684 \\
	40. &25278 &5. &0.631 &0.277 &51.578 &186.12\phn &802.816 &13.493 &2.721 \\
	41. &56997 &5.304 &0.768 &0.445 &84.238 &189.294 &887.054 &13.938 &2.766 \\
	42. &86796 &5.118 &0.696 &0.269 &62.31\phn &231.691 &949.363 &14.207 &2.801 \\
	43. &76829 &4.641 &0.523 &0.127 &29.467 &231.952 &978.831 &14.334 &2.821 \\
	44. &17651 &4.214 &0.431 &0.061 &15.664 &257.2\phn\phn &994.495 &14.395 &2.831 \\
	45. &86736 &4.864 &0.585 &0.161 &41.604 &258.559 &$1.036\times10^3$ &14.556 &2.855 \\
	46. &3909 &5.169 &0.679 &0.261 &67.586 &258.59\phn &$1.104\times10^3$ &14.817 &2.888 \\
	47. &80337 &5.366 &0.765 &0.338 &93.334 &276.502 &$1.197\times10^3$ &15.154 &2.927 \\
	48. &61174 &4.29 &0.444 &0.063 &17.475 &277.771 &$1.214\times10^3$ &15.217 &2.937 \\
	49. &4151 &4.786 &0.575 &0.126 &36.84\phn &291.855 &$1.251\times10^3$ &15.344 &2.955 \\
	50. &32480 &5.242 &0.711 &0.232 &76.179 &328.408 &$1.328\times10^3$ &15.576 &2.985 \\
	51. &84862 &5.384 &0.768 &0.291 &96.115 &330.28\phn &$1.424\times10^3$ &15.867 &3.02\phn \\
	52. &109422 &4.931 &0.606 &0.136 &46.253 &341.003 &$1.47\phn\times10^3$ &16.002 &3.039 \\
	53. &910 &4.892 &0.594 &0.127 &43.459 &343.028 &$1.513\times10^3$ &16.129 &3.057 \\
	54. &49081 &5.381 &0.78 &0.274 &95.563 &348.374 &$1.609\times10^3$ &16.403 &3.09\phn \\
	55. &114622 &5.57 &0.966 &0.374 &130.908 &349.855 &$1.74\phn\times10^3$ &16.777 &3.127 \\
	56. &48113 &5.083 &0.67 &0.168 &58.902 &351.091 &$1.799\times10^3$ &16.945 &3.15\phn \\
	57. &15330 &5.53 &0.826 &0.341 &122.436 &359.57\phn &$1.921\times10^3$ &17.286 &3.185 \\
	58. &40843 &5.127 &0.661 &0.174 &63.17\phn &362.362 &$1.984\times10^3$ &17.46\phn &3.207 \\
	59. &92043 &4.204 &0.432 &0.043 &15.455 &362.851 &$2.\phn\phn\phn\times10^3$ &17.503 &3.214 \\
	60. &22263 &5.484 &0.809 &0.312 &113.361 &363.154 &$2.113\times10^3$ &17.815 &3.247 \\
	61. &64792 &5.193 &0.697 &0.188 &70.241 &373.413 &$2.183\times10^3$ &18.003 &3.271 \\
	62. &79672 &5.487 &0.814 &0.29 &113.887 &392.079 &$2.297\times10^3$ &18.293 &3.302 \\
	63. &8362 &5.62 &0.909 &0.355 &142.309 &401.381 &$2.44\phn\times10^3$ &18.648 &3.336 \\
	64. &58576 &5.535 &0.866 &0.303 &123.513 &408.066 &$2.563\times10^3$ &18.951 &3.368 \\
	65. &113357 &5.448 &0.802 &0.257 &106.797 &415.353 &$2.67\phn\times10^3$ &19.208 &3.396 \\
	66. &73996 &4.922 &0.598 &0.107 &45.552 &426.348 &$2.715\times10^3$ &19.315 &3.41\phn \\
	67. &97675 &5.119 &0.669 &0.14 &62.371 &445.208 &$2.778\times10^3$ &19.455 &3.428 \\
	68. &29800 &5.037 &0.63 &0.122 &54.657 &449.05\phn &$2.832\times10^3$ &19.576 &3.443 \\
	69. &12653 &5.396 &0.763 &0.211 &97.937 &463.837 &$2.93\phn\times10^3$ &19.787 &3.467 \\
	70. &71957 &3.867 &0.366 &0.019 &9.654 &495.27\phn &$2.94\phn\times10^3$ &19.807 &3.47\phn \\
	71. &40702 &4.057 &0.4 &0.025 &12.537 &505.255 &$2.953\times10^3$ &19.832 &3.474 \\
	72. &59072 &4.137 &0.412 &0.027 &14.037 &519.638 &$2.967\times10^3$ &19.859 &3.477 \\
	73. &99825 &5.73 &0.983 &0.327 &171.42\phn &524.01\phn &$3.138\times10^3$ &20.186 &3.509 \\
	74. &42438 &5.63 &0.859 &0.27 &144.783 &536.696 &$3.283\times10^3$ &20.456 &3.537 \\
	75. &107649 &5.571 &0.833 &0.243 &131.021 &539.538 &$3.414\times10^3$ &20.699 &3.563 \\
	76. &32439 &5.442 &0.77 &0.194 &105.837 &544.254 &$3.52\phn\times10^3$ &20.893 &3.585 \\
	77. &50564 &4.779 &0.563 &0.064 &36.459 &573.128 &$3.556\times10^3$ &20.957 &3.593 \\
	78. &97295 &5. &0.624 &0.09 &51.552 &574.615 &$3.608\times10^3$ &21.046 &3.604 \\
	79. &75181 &5.65 &0.872 &0.258 &149.742 &579.626 &$3.757\times10^3$ &21.305 &3.631 \\
	80. &19893 &4.255 &0.434 &0.029 &16.628 &581.566 &$3.774\times10^3$ &21.333 &3.635 \\
	81. &34065 &5.575 &0.839 &0.226 &131.973 &583.891 &$3.906\times10^3$ &21.559 &3.659 \\
	82. &35136 &5.545 &0.818 &0.215 &125.535 &584.373 &$4.032\times10^3$ &21.774 &3.682 \\
	83. &7978 &5.518 &0.802 &0.205 &120.086 &586.575 &$4.152\times10^3$ &21.979 &3.704 \\
	84. &38908 &5.591 &0.834 &0.229 &135.555 &593.028 &$4.287\times10^3$ &22.207 &3.727 \\
	85. &81300 &5.765 &0.991 &0.305 &182.038 &596.193 &$4.469\times10^3$ &22.513 &3.756 \\
	86. &18859 &5.381 &0.749 &0.158 &95.622 &605.974 &$4.565\times10^3$ &22.671 &3.773 \\
	87. &73184 &5.712 &1.089 &0.272 &166.307 &611.836 &$4.731\times10^3$ &22.942 &3.799 \\
	88. &25110 &5.086 &0.651 &0.095 &59.12\phn &622.899 &$4.79\phn\times10^3$ &23.037 &3.811 \\
	89. &16245 &4.709 &0.541 &0.052 &32.707 &633.375 &$4.823\times10^3$ &23.089 &3.817 \\
	90. &34834 &4.478 &0.481 &0.035 &23.059 &656.467 &$4.846\times10^3$ &23.124 &3.822 \\
	91. &36439 &5.356 &0.732 &0.139 &91.71\phn &661.85\phn &$4.938\times10^3$ &23.263 &3.837 \\
	92. &110649 &5.315 &0.756 &0.126 &85.835 &682.18\phn &$5.024\times10^3$ &23.388 &3.852 \\
	93. &51523 &4.887 &0.593 &0.061 &43.118 &708.257 &$5.067\times10^3$ &23.449 &3.859 \\
	94. &29650 &5.2 &0.68 &0.099 &71.099 &721.585 &$5.138\times10^3$ &23.548 &3.87\phn \\
	95. &86486 &4.758 &0.552 &0.049 &35.269 &723.745 &$5.173\times10^3$ &23.597 &3.876 \\
	96. &86614 &4.562 &0.506 &0.033 &26.166 &793.268 &$5.199\times10^3$ &23.63\phn &3.881 \\
	97. &98767 &5.73 &0.935 &0.206 &171.563 &832.497 &$5.371\times10^3$ &23.836 &3.902 \\
	98. &91438 &5.857 &0.968 &0.255 &212.773 &835.214 &$5.584\times10^3$ &24.09\phn &3.926 \\
	99. &3583 &5.795 &0.932 &0.228 &191.347 &840.454 &$5.775\times10^3$ &24.318 &3.948 \\
	100. &26394 &5.65 &0.864 &0.175 &149.742 &857.238 &$5.925\times10^3$ &24.493 &3.966 \\
	101. &3093 &5.88 &1.041 &0.252 &221.312 &879.089 &$6.146\times10^3$ &24.744 &3.99\phn \\
	102. &950 &5.24 &0.693 &0.086 &75.872 &883.632 &$6.222\times10^3$ &24.83\phn &4.\phn\phn\phn \\
	103. &33277 &5.75 &0.905 &0.186 &177.37\phn &953.291 &$6.399\times10^3$ &25.016 &4.019 \\
	104. &56452 &5.962 &1.067 &0.259 &254.666 &984.962 &$6.654\times10^3$ &25.275 &4.043 \\
	105. &114924 &5.58 &0.826 &0.132 &133.067 &$1.008\times10^3$ &$6.787\times10^3$ &25.407 &4.057 \\
	106. &98819 &5.787 &0.922 &0.173 &188.837 &$1.091\times10^3$ &$6.976\times10^3$ &25.58\phn &4.074 \\
	107. &111449 &5.21 &0.684 &0.065 &72.281 &$1.107\times10^3$ &$7.048\times10^3$ &25.645 &4.082 \\
	108. &40693 &5.947 &1.046 &0.224 &248.394 &$1.108\times10^3$ &$7.296\times10^3$ &25.87\phn &4.103 \\
	109. &29860 &5.7 &0.886 &0.147 &162.931 &$1.111\times10^3$ &$7.459\times10^3$ &26.016 &4.118 \\
	110. &114948 &5.651 &0.847 &0.124 &149.972 &$1.208\times10^3$ &$7.609\times10^3$ &26.141 &4.131 \\
	111. &88175 &4.611 &0.515 &0.023 &28.171 &$1.214\times10^3$ &$7.638\times10^3$ &26.164 &4.134 \\
	112. &43587 &5.951 &1.093 &0.201 &250.024 &$1.242\times10^3$ &$7.888\times10^3$ &26.365 &4.154 \\
	113. &89348 &4.994 &0.617 &0.041 &51.079 &$1.246\times10^3$ &$7.939\times10^3$ &26.406 &4.158 \\
	114. &98470 &5.646 &0.842 &0.118 &148.685 &$1.261\times10^3$ &$8.087\times10^3$ &26.524 &4.171 \\
	115. &72567 &5.87 &0.954 &0.159 &217.466 &$1.365\times10^3$ &$8.305\times10^3$ &26.683 &4.187 \\
	116. &39780 &5.3 &0.743 &0.061 &83.688 &$1.366\times10^3$ &$8.389\times10^3$ &26.745 &4.194 \\
	117. &27435 &5.97 &1.01 &0.186 &258.268 &$1.389\times10^3$ &$8.647\times10^3$ &26.931 &4.212 
\enddata
\end{deluxetable}

\begin{deluxetable}{llllll}
\tablewidth{0pt}
\tablecaption{Trade Space of $\Delta mag_0$ for 1 Year of Exposure Time}
\tablehead{\colhead{Quantity} &\colhead{Case A} &\colhead{Case B} &\colhead{Case C} &\colhead{Case D} &\colhead{Units}}
\startdata
$y$ &\phn8/3.5 &\phn1 &\phn8/3.5 &\phn1 &\multicolumn{1}{c}{\nodata}\\
Optimized~~ &no &no &yes &yes &\multicolumn{1}{c}{\nodata}\\
$\Delta mag_0$ &25.00 &25.00 &25.45 &26.10 &delta magnitudes\\
$<$$\Sigma n_\mathrm{p}$$>$ &26.9$\pm$4.2 &34.5$\pm$5.0 &29.1$\pm$4.1 &63.3$\pm$6.0 &planets discovered\\
$n_\mathrm{s}$ &\llap{1}17 ~~ &\llap{3}24 &90~~ &\llap{1}76&stars observed\\
$\delta \sigma^2$ &(0.0092)$^2$ &(0.0140)$^2$ &(0.0075)$^2$ &(0.0084)$^2$ &nm$^2$\\
\enddata
\tablecomments{The value of $<$$\Sigma n_\mathrm{p}$$>$ assumes that every star has one planet drawn from the habitable, Earth-like
population. The results for $\delta\sigma^2$  are derived from $\Delta mag_0$ using equation~(26).}
\end{deluxetable}

\end{document}